\documentstyle[prl,twocolumn,aps,floats,epsfig]{revtex}

\begin{document}

\draft
\twocolumn[\hsize\textwidth\columnwidth\hsize\csname
@twocolumnfalse\endcsname

\title{$c$\,-axis phase coherence and spin fluctuations in cuprates}
\author{A. Mourachkine}
\address{Universit\'{e} Libre de Bruxelles, CP-232, 
Blvd du Triomphe, B-1050 Brussels, Belgium}

\date{Received 5 June 2000}
\maketitle

\begin{abstract}
{\bf Abstract.} - There is a consensus that superconductivity (SC) in 
cuprates is two-dimensional. It is widely believed that the long-range 
phase coherence appears at $T_{c}$ due to the Josephson coupling between 
SC CuO$_{2}$ (bi-, tri-, ...)layers. Recent $T_{c}$ and resistivity 
measurements in Tl$_{2}$Ba$_{2}$CaCu$_{2}$O$_{8}$ as a function of 
applied pressure (Salvetat J.-P. {\it et al.}, {\it Europhys. Lett.}, {\bf 52} 
(2000) 584) show that the 
interlayer Josephson-coupling mechanism does not fit the data. Here we 
analyze data obtained in Andreev reflection, neutron scattering, microwave, 
muon spin relaxation, tunneling and resistivity measurements performed on 
different cuprates, mainly, on YBa$_{2}$Cu$_{3}$O$_{6+x}$, 
Bi$_{2}$Sr$_{2}$CaCu$_{2}$O$_{8+x}$ and La$_{2-x}$Sr$_{x}$CuO$_{4+y}$. 
The analysis of the data shows that the long-range phase coherence in the 
cuprates intimately relates to antiferromagnetic interactions along the 
$c$ axis. At the same time, it seems that the in-plane mechanism of the SC 
has no or little relations to the magnetic interactions along the $c$ axis.
\end{abstract}


]

\section{Introduction} 

Superconductivity (SC) and magnetism were earlier considered as mutually 
exclusive phenomena. Recent research revealed a rich variety of 
extraordinary SC and magnetic states and phenomena in novel materials 
that are due to the interaction between SC and magnetism \cite{Maple}. 
The coexistence of SC and long-range antiferromagnetic (AF) order was first 
discovered in 
$R$Mo$_{6}$Se$_{8}$ ($R$ = Gd, Tb and Er), $R$Rh$_{4}$B$_{4}$ ($R$ = Nd, 
Sm and Tm), and $R$Mo$_{6}$S$_{8}$ ($R$ = Gd, Tb, Dy and Er) \cite{Maple}. 
Later, coexistence of SC and AF order was found in U-based heavy fermions 
(UPt$_{3}$, URu$_{2}$Si$_{2}$, UNi$_{2}$Al$_{3}$, UPd$_{2}$Al$_{3}$, 
U$_{6}$Co, and U$_{6}$Fe), in heavy fermions $R$Rh$_{2}$Si$_{2}$ ($R$ = 
La and Y), Cr$_{1-x}$Re$_{x}$, CeRu$_{2}$, in borocarbides 
$R$Ni$_{2}$B$_{2}$C ($R$ = Tm, Er, Ho, Dy), in organic SCs 
\cite{Maple,Gabi1} and in the new heavy fermion 
CeRh$_{1-x}$Ir$_{x}$In$_{5}$ \cite{LosAmos}. In 
CeRh$_{0.5}$Ir$_{0.5}$In$_{5}$, the bulk SC coexists {\em microscopically} 
with small-moment magnetism ($\leq$ 0.1$\mu$$_{B}$) \cite{LosAmos}. 
In all other heavy fermions, there are strong AF correlations present in 
the SC state \cite{Maple,Gabi1,LosAmos}. Another class of materials in 
which SC and AF coexist are copper oxides (cuprates) such as 
YBa$_{2}$Cu$_{3}$O$_{6+x}$ (YBCO) and 
La$_{2-x}$Sr$_{x}$CuO$_{4+y}$ (LSCO) compounds 
\cite{Maple}. Coexistence of SC and ferromagnetic (FM) order is found in the 
heavy fermion UGe$_{2}$ \cite{ferro}, and in Ru-based materials, for example, 
in RuSr$_{2}$GdCu$_{2}$O$_{8}$ \cite{Ru}.

In SC heavy-fermion systems, spin fluctuation (electron-electron 
interactions) are believed to mediate the electron pairing that leads to 
SC \cite{Varma}. For the heavy fermions CeIn$_{3}$, CePd$_{2}$Si$_{2}$ 
\cite{Martur}, UPd$_{2}$Al$_{3}$ \cite{Hunt} and UGe$_{2}$ \cite{ferro}, 
there is an indirect evidence for spin-fluctuation mechanism of SC. This 
intimate relationship between the SC and magnetism also appears to be 
central to cuprates \cite{Maple} which inherited magnetic 
properties from their parent compounds, AF Mott insulators. Many 
theoretical studies suggest that the SC in cuprates is mediated via the 
exchange of AF spin fluctuations \cite{Pine}. 

There is a consensus that, in the underdoped regime of cuprates,
there are two energy scales \cite{Deu}: the pairing energy scale, 
$\Delta$$_{p}$, and the phase-coherence scale, $\Delta$$_{c}$, observed 
experimentally \cite{Deu,AMour3}. The two energy scales have different 
dependences on hole concentration, $p$, in CuO$_{2}$ planes (see fig. 3): 
$\Delta$$_{p}$ increases linearly with decrease of hole concentration, 
whereas $\Delta$$_{c}$ has approximately the parabolic dependence on 
$p$ and scales with $T_{c}$ as 2$\Delta$$_{c}$ $\simeq$ 5.4$k_{B}T_{c}$ 
\cite{Deu}.

There is a consensus that the SC in cuprates is two-dimensional (2D). 
It is widely believed that the long-range phase coherence occurs at $T_{c}$ 
due to the Josephson coupling between SC CuO$_{2}$ (bi-, tri-, ...)layers. 
Recent measurements of the in-plane ($\rho$$_{ab}$) and out-of-plane 
($\rho$$_{c}$) resistivities in Tl$_{2}$Ba$_{2}$CaCu$_{2}$O$_{8}$ as a 
function of applied pressure show that $\rho$$_{c}(T)$ shifts smoothly down 
with increase of pressure, however, $T_{c}$ first increases and then 
{\em decreases} \cite{Tl2212}. This result can not be explained 
by the interlayer Josephson-coupling mechanism. The authors conclude 
\cite {Tl2212}: ``Any model that associates high-$T_{c}$ with the interplane 
Josephson coupling should therefore be revisited.'' In this paper, we analyze 
data obtained in Andreev reflection, inelastic neutron 
scattering (INS), microwave, muon spin relaxation ($\mu$SR), tunneling and 
resistivity measurements performed on different cuprates, mainly, on YBCO, 
Bi$_{2}$Sr$_{2}$CaCu$_{2}$O$_{8+x}$ (Bi2212) and LSCO. 
The analysis of the data shows that the long-range phase coherence in the 
cuprates intimately relates to AF interactions along the $c$ axis 
\cite{note3}. At the same time, it seems that the in-plane mechanism of the 
SC has no or little relations to the magnetic interactions along the $c$ 
axis. We also analyze data measured in heavy fermions UPt$_{3}$, 
UPd$_{2}$Al$_{3}$ and CeIrIn$_{5}$, and in some layered non-SC compounds 
with FM correlations.

\section{In-plane and $c$\,-axis tunneling in Bi2212} 

Tunneling spectroscopy
is an unique probe of SC state in that it can, in principle, reveal the 
quasiparticle density of states directly with high energy resolution.
Figure 1 shows the temperature dependence of in-plane tunneling 
quasiparticle peaks, measured in slightly overdoped Bi2212 single 
crystals \cite{Ekino1,Matsuda,AMour1,Oda}. 
Tunneling measurements performed on slightly underdoped Bi2212 single 
crystals show similar temperature dependence \cite{Miyakawa}. So, the 
temperature dependence of in-plane quasiparticle peaks shown in fig. 1 can 
be considered as a typical one. In fig. 1, we present also the temperature 
dependence of $c$\,-axis quasiparticle peaks measured in slightly 
overdoped Bi2212 mesas \cite{Yurgens}. Measurements performed on 
micron-size mesas present {\em intrinsic} properties of the material. 
In fig. 1, surprisingly, the temperature dependences along the $ab$ planes 
and along the $c$ axis are {\em different}. It is a clear hallmark of 
coexistence of two different SC mechanisms in Bi2212: in-plane and along 
the $c$ axis.

\section{LSCO} 

In LSCO, there is an evidence that the SC intimately relates to 
the establishment of AF order along the $c$ axis. 

Recent $\mu$SR 
measurements performed on non-SC Eu-doped LSCO having different hole 
concentrations show that the SC phase of pure LSCO is replaced in Eu-doped 
LSCO by the second AF phase (see fig. 4 in Ref.\cite{LSCO}). Thus, the data 
show that it is possible to switch the entire hole concentration dependent 
phase diagram from SC to AF. It is a clear hallmark that the SC in LSCO 
intimately relates to the formation of AF order. We return to this 
important result later.

We turn now to the analysis of resistivity data measured in non-SC 
2D layered compounds with AF or FM correlations. The data clearly 
show that, in all these layered compounds, the out-of-plane resistivity, 
$\rho$$_{c}$, has drastic changes either at N\'{e}el temperature, $T_{N}$, 
or Curie temperature, $T_{C}$, whereas the in-plane resistivity,
\begin{figure}[t]
\leftskip-10pt
\epsfxsize=0.9\columnwidth
\centerline{\epsffile{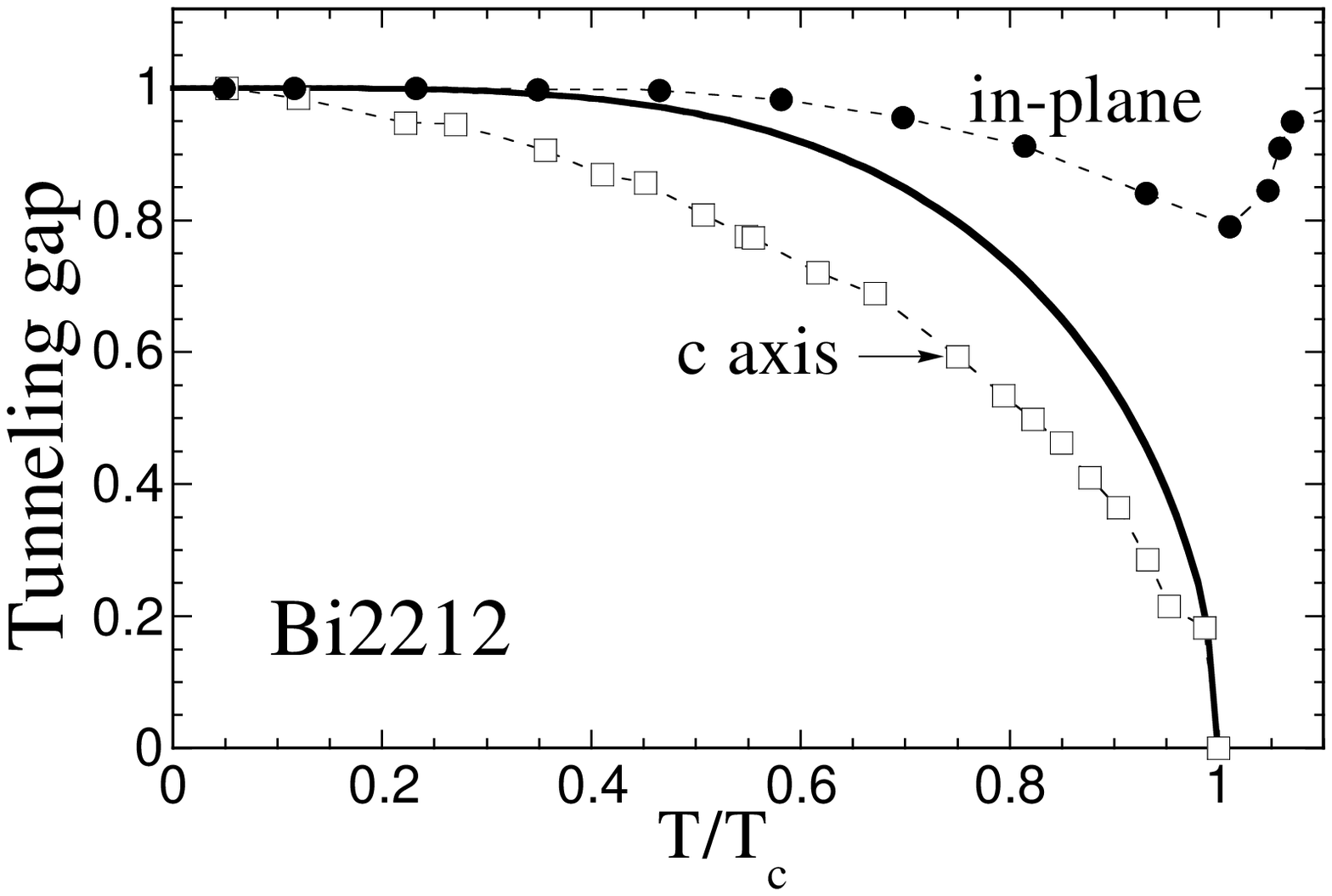}}
\vspace{2mm}
\caption{Temperature dependences of in-plane (dots) [14-17] 
and $c$\,-axis  tunneling quasiparticle peaks (squares) [19] in slightly 
overdoped Bi2212 single crystals, $\Delta (T)/ \Delta (T_{min})$. 
The BCS temperature dependence is shown by the solid line. 
The dashed lines are guides  to the eye.}
\label{fig1}
\end{figure} 
$\rho$$_{ab}$, passes through $T_{N}$ or $T_{C}$ smoothly. 

We start with YBCO. In AF undoped YBCO (x = 0.35; 0.33; and 0.32) having 
$T_{N}$ $\simeq$ 80 K, 160 K, and 210 K, respectively, $\rho$$_{c}$ 
shows sharp increase, by about 2 orders of magnitude, upon cooling through 
$T_{N}$ \cite{Lavrov1}. At the same time, $\rho$$_{ab}$ changes at 
$T_{N}$ smoothly. The same effect has been observed in LuBCO (x = 0.34) 
\cite{Lavrov2}. So, the N\'{e}el ordering in undoped YBCO has 
remarkably different impact on the electron transport within CuO$_{2}$ 
planes and between them. The authors conclude \cite{Lavrov1}: 
``The N\'{e}el temperature actually corresponds to the establishment of AF 
order along $c$ axis.''

We now discuss FM layered compounds. The structure of the FM compound 
Bi$_{2}$Sr$_{3}$Co$_{2}$O$_{9}$ (BSCoO) is similar to the structure of 
Bi2212, where CoO$_{2}$ planes are analogous with CuO$_{2}$ planes in 
Bi2212 \cite{Co}. BSCoO becomes FM at $T_{C}$ = 3.2 K. The resistivity 
data show that, at $T_{C}$, there is a cusp in $\rho$$_{c}$ but 
$\rho$$_{ab}$ changes smoothly. The authors conclude that the long-range 
magnetic order in BSCoO develops at $T_{C}$ along the $c$ axis \cite{Co}. 
In layered manganite La$_{1.4}$Sr$_{1.6}$Mn$_{2}$O$_{7}$ (LSMO) which 
is composed of the MnO$_{2}$ bilayers becomes FM at $T_{C}$ = 90 K 
\cite{MnO}. The resistivity data show that, at $T_{C}$, there are drastic 
changes in $\rho$$_{c}$ (a few orders of magnitude), but very small 
changes in $\rho$$_{ab}$. They conclude that, in LSMO, the long-range 
magnetic order develops at $T_{C}$ = 90 K along the $c$ axis \cite{MnO}.

Neutron-scattering measurements performed on the heavy fermion 
URu$_{2}$Si$_{2}$ ($T_{c}$ = 1.2 K) show that the AF order develops at 
$T_{N}$ = 17.5 K along the $c$ axis \cite{heavferm}. So, it seems that, in 
all layered compounds, the long-range AF or FM order develops at $T_{N}$ 
or $T_{C}$ along the $c$ axis (the {\em in-plane} magnetic correlations 
exist above $T_{N}$ and $T_{C}$ \cite{MnO}).

We now return to the analysis of the phase diagram of non-SC Eu-doped 
LSCO, where the SC phase of pure LSCO is replaced by the second AF 
phase \cite{LSCO}. The conclusion made in the previous paragraph 
signifies that either the main AF phase of Eu-doped LSCO or the second 
AF phase develops along the $c$ axis. Thus, the SC phase of pure LSCO is 
replaced in Eu-doped LSCO by the AF phase which develops along the $c$ 
axis. Consequently, the SC in LSCO intimately relates to the 
establishment of the long-range AF order along the $c$ axis.

In heavy fermion CeIrIn$_{5}$, $\mu$SR measurements discovered the 
onset of a small magnetic field ($\sim$ 0.4 Gauss) which sets exactly at 
$T_{c}$ \cite{LosAmos}. In YBCO (x = 0.6), recent INS measurements 
identified small magnetic moments directed along the $c$ axis, which 
increase in strength at and below $T_{c}$ \cite{Mook}.

\section{YBCO, Bi2212 and LSCO}
 
We now compare coherence SC and magnetic characteristics of YBCO, 
Bi2212 and LSCO. The comparison shows that the magnetic and coherence 
SC characteristics have similar temperature dependencies, and, at 
different dopings, their magnitudes are proportional to each other (and 
proportional to $T_{c}$). Thus, the coherence SC and magnetic properties
\begin{figure}[t]
\leftskip-10pt
\epsfxsize=0.8\columnwidth
\centerline{\epsffile{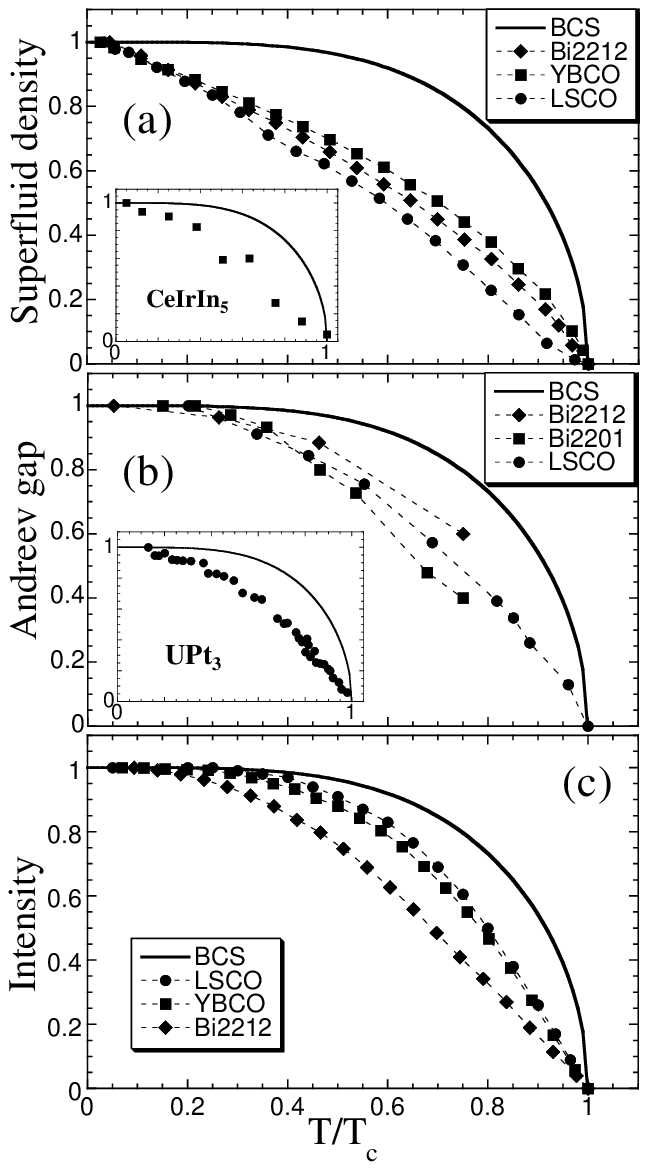}}
\end{figure}  
of cuprates intimately relate to each other.

First, we describe the magnetic properties of cuprates. The low energy 
magnetic excitations in LSCO cuprate have been extensively studied, and 
the observed spin fluctuations are characterized by wave vector which 
is incommensurate with the lattice \cite{Lee}. These modulated spin 
fluctuations in LSCO persist in both normal and SC states. The spin 
dynamics in YBCO and Bi2212 studied by INS exhibit below $T_{c}$ a 
sharp commensurate resonance peak which appears at well defined 
energy $E_{r}$ \cite{i1,i2,i3,i4,i5,i6,i7,i8,i9,i10,i11}. 
Incommensurability in YBCO has been also reported \cite{i8}, and 
it is consistent with that in LSCO of the same hole doping, but, in YBCO, 
it occurs in the SC state. Now it is clear that the incommensurability 
and the commensurate resonance are inseparable parts of the general 
features of the spin dynamics in YBCO at all doping levels \cite{Mook2}.
Thus, there is a clear evidence of coexistence of AF order and SC below 
$T_{c}$, at least, in LSCO and YBCO. 

We now compare coherence SC and  magnetic characteristics of the
\begin{figure}[t]
\caption{(a) Temperature dependence of the superfluid density in near 
optimally doped single crystals of Bi2212 ($T_{c}$ = 93 K) [40] 
and YBCO ($T_{c}$ = 93 K) [41], and in an overdoped LSCO (x = 0.2) 
single crystal ($T_{c}$ = 36 K) [42]. The superfluid density is 
proportional to 1/$\lambda ^{2}(T)$, where $\lambda (T)$ is the 
magnetic penetration depth. Inset: temperature dependence of the 
superfluid density in heavy fermion CeIrIn$_{5}$ 
($T_{c}$ = 0.4 K) [3] (axis parameters as main plot). 
(b) Temperature dependence of Andreev-reflection gap, 
$\Delta$$(T)$/$\Delta$($T_{min}$), in an overdoped Bi2212 thin film 
($T_{c}$ = 80 K) [43], and in overdoped single crystals of 
Bi2201 ($T_{c}$ = 29 K) [44] and LSCO (x = 0.2) ($T_{c}$ = 28 K) 
[45]. Inset: temperature dependence of the Andreev gap 
in heavy fermion UPt$_{3}$ ($T_{c}$ $\sim$ 440 mK) [48] (axis 
parameters as main plot). (c) Temperature dependence of the peak 
intensity of the incommensurate elastic scattering in LSCO (x = 0) 
($T_{c}$ = 42 K) [27] and the intensity of the magnetic resonance 
peak measured by INS in near optimally doped Bi2212 ($T_{c}$ = 91 K) 
[28] and YBCO ($T_{c}$ = 92.5 K) [33]. The neutron-scattering 
data are average, the real data have the vertical error of the order 
of $\pm$10\% [27,28,33]. The BCS temperature dependence is 
shown by the thick solid line. The dashed lines are guides to the eye.}
\label{fig2}
\end{figure} 
cuprates. Figure 2(a) shows the temperature dependences of the 
superfluid density in near optimally doped single crystals of Bi2212 
\cite{rho1} and YBCO \cite{rho2}, and in an overdoped LSCO (x = 0.2) 
single crystal \cite{rho3}, measured by microwave, $\mu$SR and 
ac-susceptibility techniques, respectively. The superfluid density is 
proportional to 1/$\lambda$$^{2}(T)$, where $\lambda$$(T)$ is the 
magnetic penetration depth. Figure 2(b) shows the temperature 
dependences of Andreev-reflection gap measured in an overdoped Bi2212 
thin film \cite{andreev1}, and in overdoped single crystals of Bi2201 
\cite{andreev2} and LSCO (x = 0.2) \cite{andreev3}. It is important to 
emphasize that Andreev reflections are exclusively sensitive to 
coherence properties of the condensate. In figs 2(a) and 2(b), one can 
see that there is a good agreement among temperature dependences of 
coherence SC characteristics of different cuprates. 

Figure 2(c) shows the temperature dependences of the peak intensity of 
the incommensurate elastic scattering in LSCO (x = 0) \cite{Lee} and the 
intensity of the commensurate resonance peak measured by INS in near 
optimally doped Bi2212 \cite{i1} and YBCO \cite{i6}. 

In fig. 2, all temperature dependences of coherence SC and magnetic 
characteristics exhibit below $T_{c}$ a striking similarity. Since 
all temperature dependences shown in fig. 2 are similar to the 
temperature dependence of $c$\,-axis quasiparticle peaks in Bi2212, 
shown in fig. 1, and different from the temperature dependence of 
in-plane quasiparticle peaks, shown also in fig. 1, it is reasonable to 
assume that the coherence SC and magnetic properties of the cuprates 
intimately relate to each other along the $c$ axis. Moreover, in YBCO 
(x = 0.6), recent INS measurements found small magnetic moments directed 
along the $c$ axis, which increase in strength at and below $T_{c}$ 
\cite{Mook}. At the same time, as one can see in figs 1 and 2, the in-plane 
mechanism of the SC in the cuprates has no or little relations to the
magnetic interactions along the $c$ axis. 

Figure 3 shows the phase diagram of cuprates \cite{Deu}. As noted earlier,
$\Delta$$_{p}$ is the pairing energy scale, and $\Delta$$_{c}$ is the 
energy scale which is responsible for the long-range phase coherence
and proportional to $T_{c}$ as 2$\Delta$$_{c}$ $\simeq$ 5.4$k_{B}T_{c}$ 
\cite{Deu}. At different doping levels, Andreev-reflection data coincide 
with $\Delta$$_{c}$ \cite{Deu,AMour3,andreev3}. In fig.3, we present also
the energy position of the magnetic resonance peak, 
$E_{r}$, in Bi2212 \cite{i1,i2} and in YBCO 
\cite{i3,i4,i5,i6,i7,i8,i9,i10,i11} as a function of doping.  In fig.3, one 
can see that, at different dopings, $E_{r}$ is proportional to 
$\Delta$$_{c}$ as $E_{r}$ $\simeq$ 2$\Delta$$_{c}$. 
This correlation suggests that spin excitations are responsible for 
establishing the phase coherence in YBCO and Bi2212 since the relation 
$E_{r}$ $\simeq$ 2$\Delta$$_{c}$ is in a good agreement with the 
theories in which SC is mediated by spin fluctuations \cite{Pine}.
Recently, it was shown that applied magnetic fields suppress the 
magnetic resonance peak in YBCO indicating that the resonance peak, 
indeed, measures the long-range phase coherence \cite{Dai}. The 
strength of coupling between spin excitations and charge carriers is 
sufficient to account for the high $T_{c}$ value in cuprates \cite{Basov}.
\begin{figure}[t]
\leftskip-10pt
\epsfxsize=0.9\columnwidth
\centerline{\epsffile{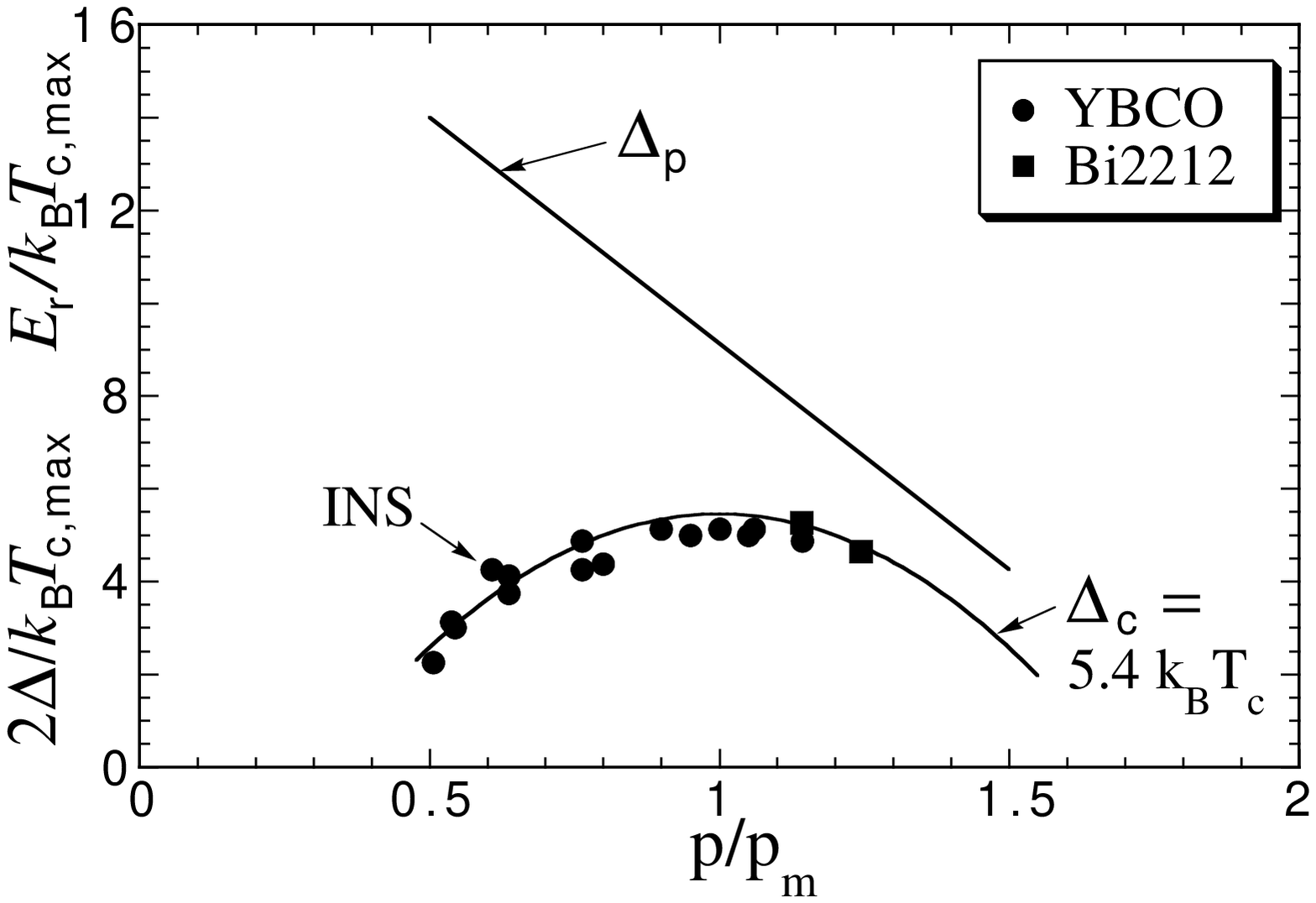}}
\vspace{2mm}
\caption{The phase diagram of cuprates (lines) [10] and the energy 
position of the magnetic resonance peak, $E_{r}$, in Bi2212 
[28,29] and in YBCO [30--38] at 
different hole concentrations ($p_{m}$ = 0.16). 
When not available, $p$ has been calculated from the relation 
$T_{c}$/$T_{c,max}$ = 1 - 82.6($p$ - 0.16)$^{2}$ [50], and we 
use $T_{c,max}$ = 93 K and $T_{c,max}$ = 95 K in the case of YBCO and 
Bi2212, respectively.}
\label{fig3}
\end{figure}

What is interesting is that all temperature dependences shown in fig.2 
are similar to the temperature dependence of the superfluid density in 
heavy fermion CeIrIn$_{5}$, measured by $\mu$SR\cite{LosAmos}, and to 
the temperature dependence of the Andreev-reflection gap in heavy 
fermion UPt$_{3}$ \cite{DeWilde}, which are shown in the insets of 
figs 2(a) and 2(b), respectively. Spin fluctuations are believed to 
mediate the electron pairing in CeIrIn$_{5}$ \cite{LosAmos} and UPt$_{3}$ 
\cite{Varma} that leads to SC. The magnetic resonance peak has not yet 
been detected in CeIrIn$_{5}$ or UPt$_{3}$, however, the magnetic 
resonance peak has been observed by INS in another heavy fermion 
UPd$_{2}$Al$_{3}$ \cite{Metoki} where spin fluctuations mediate the SC 
which coexists with the long-range AF order \cite{Hunt}. The latter facts 
point also to presence of spin-fluctuation coupling mechanism in cuprates. 

The behavior of all temperature dependences shown in fig. 2 can be 
easily understood in terms of the spin-fluctuation mechanism of SC
(electron-electron interactions): crudely speaking, they exhibit the 
squared BCS temperature dependence.

\section{Discussion} 

In spite of the unmistakable similarities among the magnetic and SC 
properties of YBCO, Bi2212 and LSCO (and some heavy fermions for 
which there is an indirect evidence for spin-fluctuation mechanism of 
SC), clearly, there is a difference between magnetic properties of LSCO 
and YBCO. If, in YBCO, $T_{c} \simeq T_{com} \simeq T_{inc}$, where 
$T_{com}$ 
($T_{inc}$) is the onset temperature of the (in-) commensurate peak(s), 
in LSCO, the situation is different. First, the commensurate peak has 
not been detected. Second, in LSCO, mainly, $T_{c} < T_{inc}$ 
(Figure 2(c) shows the case when $T_{c}$ = $T_{inc}$). So, the magnetic 
and SC properties of LSCO are similar to those of most heavy fermions: 
the commensurate peak has not been detected, and $T_{c} < T_{N}$. 
For example, in UPd$_{2}$Al$_{3}$, $T_{c}$ $\approx$ 2 K and 
$T_{N}$ = 14.5 K \cite{Hunt,Metoki}; in URu$_{2}$Si$_{2}$, 
$T_{c}$ $\approx$ 1.2 K and $T_{N}$ = 17.5 K \cite{heavferm}, and, in 
FM heavy fermion UGe$_{2}$, $T_{c} < T_{C}$ \cite{ferro}. In Bi2212, the 
situation looks more like that of YBCO, even if, the incommensurate 
peaks have not {\em yet} been detected. Apparently, it is only a question of 
time.

In fact, the differences between magnetic and SC properties of YBCO and 
LSCO (and among some heavy fermions) can be understood in terms of the 
chain of the following events: the formation of pairs at $T_{pair}$ - the
SC order parameter couples to the magnetic in-plane order parameter - 
the appearance of AF interactions along the $c$ axis at $T_{m}$ - the 
appearance of the long-range SC phase coherence. If there is no pairs, the 
SC is absent, even if, the AF order is established (like in the Eu-doped 
LSCO \cite{LSCO}). If $T_{pair} < T_{m}$, then, it reminds the situation in 
LSCO where $T_{c} < T_{inc}$. If $T_{pair} > T_{m}$, then, it reminds the 
case of YBCO where $T_{c}$ = $T_{com}$ = $T_{inc}$, and, consequently, 
$T_{c}$ = $T_{m}$. Unfortunately, this naive picture cannot explain the 
presence of the commensurate peak in INS spectra but, at 
least, shows that the differences between the magnetic and SC properties 
of LSCO and YBCO can be understood in frameworks of one picture with 
different initial parameters.

\section{Summary}

The analysis of the data obtained by different techniques in 
YBCO, Bi2212 and LSCO shows that the long-range phase coherence 
intimately relates to AF interactions along the $c$ axis. 
At the same time, it seems that the in-plane mechanism of the SC 
has no or little relations to the magnetic interactions, at least, along the 
$c$ axis. Apparently, in cuprates, the magnetic and SC order parameters 
are coupled to each other, and the phase-coherence scale, $\Delta$$_{c}$, 
has the magnetic origin (see fig. 3).

There are common features in the SC state of the heavy fermions 
CeIrIn$_{5}$, UPt$_{3}$ and UPd$_{2}$Al$_{3}$, on the one hand, and the 
cuprates, on the other hand. It is possible that, in all heavy-fermion and 
organic SCs, the long-range phase coherence is also established due to 
spin fluctuations (the pairing mechanism may be different).

\end{document}